\documentclass[aps,amssymb,amsfonts,preprint,floatfix,showpacs,superscriptaddress,10pt]{revtex4-1}

\usepackage[utf8]{inputenc}
\usepackage[french,english]{babel}
\usepackage[T1]{fontenc}
\usepackage[version=3]{mhchem} 
\usepackage{graphicx}
\usepackage{dcolumn}
\usepackage{bm}
\usepackage{xcolor}
\usepackage{subcaption}
\usepackage{amssymb}
\usepackage{amsmath}
\usepackage{siunitx} 
\usepackage{epsfig}
\usepackage[colorlinks=false, pdfborder={0 0 0}]{hyperref}
\usepackage{cleveref}

\usepackage{lipsum}

\crefname{figure}{Figure}{Figures}
\crefname{equation}{Equation}{Equations}
\crefname{table}{Table}{Tables}

\newcommand{\jk}[1]{\textcolor{black}{#1}}

\newcommand{\mJ}{\milli\joule} 
\newcommand{\nF}{\nano\F} 
\DeclareSIUnit\pxl{pixels}
\DeclareSIUnit\ppm{ppm}
\DeclareSIUnit\ppb{ppb}
\DeclareSIUnit\ppt{ppt}
\DeclareSIUnit{\liter}{$\ell$}

\graphicspath{{./figures/}}

\begin{document}
\title{Shockwave-assisted laser filament conductivity}

\author{{Elise Schubert}}
\affiliation{Group of Applied Physics, University of Geneva, 22 Chemin de Pinchat, 1211 Geneva 4, Switzerland} 
\author{{Denis Mongin}}
\affiliation{Group of Applied Physics, University of Geneva, 22 Chemin de Pinchat, 1211 Geneva 4, Switzerland}
\author{{Thomas Produit}}
\affiliation{Group of Applied Physics, University of Geneva, 22 Chemin de Pinchat, 1211 Geneva 4, Switzerland}
\author{{Guillaume~Schimmel}}
\affiliation{Group of Applied Physics, University of Geneva, 22 Chemin de Pinchat, 1211 Geneva 4, Switzerland}
\author{{J\'er\^ome Kasparian}}\email{jerome.kasparian@unige.ch}
\affiliation{Group of Applied Physics, University of Geneva, 22 Chemin de Pinchat, 1211 Geneva 4, Switzerland}
\affiliation{Institute for Environmental Sciences, University of Geneva, bd Carl Vogt  66, 1211 Geneva 4, Switzerland} 
\author{Jean-Pierre Wolf}
\affiliation{Group of Applied Physics, University of Geneva, 22 Chemin de Pinchat, 1211 Geneva 4, Switzerland}

\begin{abstract}
We investigate the influence of ultrashort laser filaments on high-voltage discharges and spark-free unloading at various repetition rates and wind conditions. For electric fields well below, close to and above the threshold for discharges, we respectively observe remote spark-free unloading, discharge suppression, and discharge guiding. These effects rely on an indirect consequence of the thermal deposition, namely the fast dilution of the ions by the shockwave triggered by the filament at each laser shot. This dilution drastically limits recombination and increases the plasma channel conductivity that can still be non-negligible after tens or hundreds of milliseconds. As a result, the charge flow per pulse is higher at  low repetition rates.

\end{abstract}

\maketitle

\section{Introduction}

The interaction of high-power lasers with electric fields has attracted much attention since the 1970's, with the aim of controlling lightning~\cite{Ball1974,UchidSYMYYKT1999,Apollonov2002} and/or triggering and guiding high-voltage discharges~\cite{koopman1971channeling,MikiAS1993}. These early experiments implied laser pulses carrying Joules to kilojoules, and essentially relied on the local heating of the air at the laser focus. Such heating increases collisional ionization and reduces electron losses due to attachment and electron-ion recombination~\cite{Bazelyan2000}. It also decreases the air density,  favoring the electrical breakdown as described by Paschen's law~\cite{tirumala2010analytical}. 
However, the high density of plasma produced by those nanosecond high-power pulses is opaque to the laser light, which leads to a discontinuous chain of plasma beads. On the other hand, the effect drops for looser beam focusing as the incident intensity is much lower. 

The advent of ultrashort laser pulses simultaneously allowed to reach high peak powers at much lower pulse energy, and to generate plasma channels with moderate plasma density (typically \SIrange[range-phrase=--]{e20}{e23}{\m\tothe{-3}}) extending over tens to hundreds of meters~\cite{la1999filamentation}. Ultrashort laser filaments~\cite{ChinHLLTABKKS2005,berge2007ultrashort,couairon2007femtosecond} can trigger (with breakdown threshold reduction by typically \SI{30}{\%}) and guide high-voltage discharges both in the near-infrared~\cite{ComtoCDGJJKFMMPRVCMPBG2000,LaCCDGJJKMMPRVPCM2000,rodriguez2002triggering,gordon2003streamerless} and in the ultraviolet~\cite{zhao1995femtosecond,Bazelyan2000,rambo2001high,IoninKLSSSSUZ2012}.
Laser filaments have also been used as a probe to investigate the dynamics of high-voltage discharges \cite{sugiyama2009laser,sugiyama2010submicrosecond}, streamers~\cite{wang2015direct} and leaders~\cite{eto2012quenching,schmitt2015importance}. 

Most experiments have been performed with Marx generators, that produce high-voltage pulses with a temporal shape representative of a naturally developing lightning strike. However, the electric field  before the lightning initiation is close to DC. In this regime, new behaviors like discharge suppression and remote spark-free unloading~\cite{schubert2015remote,Schubert2017} or an increase of the plasma lifetime in a coupled AC-DC configuration~\cite{theberge2016laser,Point2016} have been observed.

Here, we investigate the influence of the laser repetition rate and wind on the discharge suppression and guiding, as well as of spark-free unloading. We evidence a plasma stabilization mechanism relying on the laser-induced shockwave~\cite{Lahav2014,jhajj2014demonstration} and the resulting dilution of the ion density. This stabilization allows the filaments to keep a measurable, slowly decaying conductivity hundreds of milliseconds to a few seconds after the laser pulse. It increases the estimated filament conductivity over a duty cycle, and drastically reduces its dependence on the initial plasma density.

\section{Experimental Setup}

\begin{figure}[tb!]
\centering
\includegraphics[width=0.7\columnwidth, keepaspectratio]{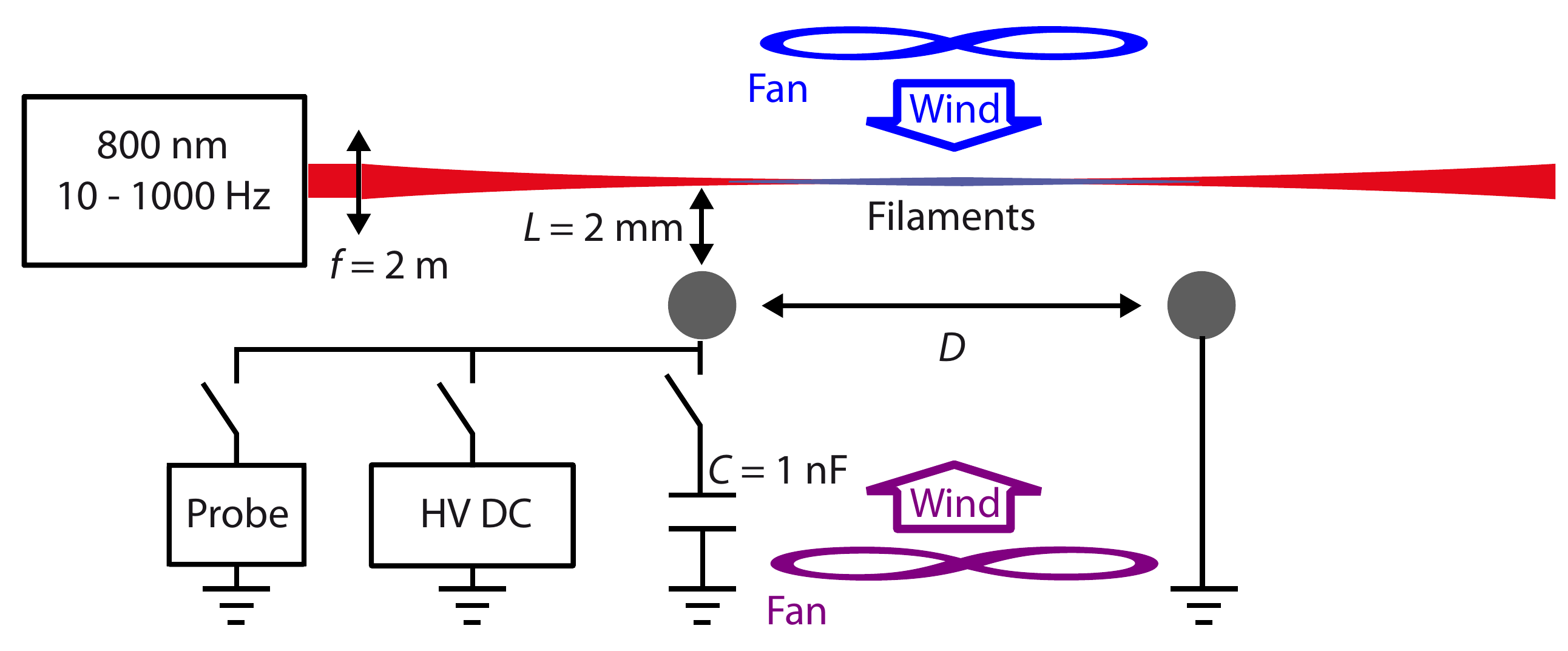} 	
\caption{Experimental setup. The DC high-voltage generator delivers \SI{14}{kV}, or \SI{100}{kV}, \SI{200}{\micro\A}. The distance between the electrodes can be varied between $D = \SI{4}{\cm}$ and \SI{16}{\cm}. Wind can be blown at \SI{\sim 2}{\m\per\s}, from either side, as well as from the top of the setup and longitudinally along the laser beam.}
\label{fig:hvcapa_schema}
\label{fig:hvrep_setup}
\end{figure}

The experimental setup is depicted in \cref{fig:hvcapa_schema}. Two spherical electrodes of \SI{1.2}{\cm} diameter were placed at a distance $D=\SIrange[range-phrase=-]{4}{16}{cm}$ from each other. One electrode was grounded, while the other one was connected to a high voltage generator. 
\SI{20}{\cm}-long filaments, covering the gap between the electrodes, were produced at a distance $L=\SI{2}{mm}$ from the electrodes by a Ti:Sapphire chirped pulse amplification system delivering pulses at a repetition rate of \SIrange{10}{1000}{\Hz}, slightly focused by an $f= \SI{2}{\m}$ lens.

Wind could be blown by a fan in the filament--electrode plane, either from the filament or from the electrode sides, as well as along the beam axis and from the top of the setup. The wind speed of  \SI{\sim 2}{\m \per \s} ensured that even at \SI{1}{kHz} repetition rate each filament sweeps a fresh air parcel, as the air motion reaches several millimeters during the \SI{1}{ms} time interval between two pulses. This displacement is typically 10 times larger than the filament diameter of \SI{100}{\micro\m}.

All experiments have been performed at room temperature ($T = \SI{20}{\celsius}$), at a relative humidity around \SI{30}{\%}. The corresponding background resistivity of the air is about \SI{3e14}{\ohm\m}~\cite{Pawar2009}.

This setup was used in two distinct regimes. 
Spark-free unloading was investigated with electrodes at a distance of $D=\SI{16}{cm}$ from each other. A $C=\SI{1}{\nF}$ capacitor was attached to the high-voltage electrode. The capacitor was initially charged under \SIrange{10}{14}{\kV}, and then disconnected from the generator. The voltage on the capacitor was periodically probed by a \numrange[range-phrase=:]{1}{4000} voltage divider of \SI{300}{\giga\ohm} impedance and a follower amplifier (\textsc{TL1169}). Without laser, the capacitor kept its charge over at least \SI{20}{hours}.
The laser producing the filaments delivered \SI{80}{\fs} pulses of \SI{14.5}{\mJ}, with an initial beam diameter of \SI{4}{cm}. The unloading is characterized by the decay time \jk{$\tau = R_\textrm{eff}C$}, obtained from an exponential fit \jk{of the voltage evolution, where $R_\textrm{eff}$ is the effective  resistance of the plasma channel~\cite{mongin2016conductivity}, from which we calculate the effective conductance per pulse}. 

Spark guiding and suppression were studied with a DC high-voltage generator delivering up to  \SI{200}{\uA} under \SI{100}{\kV}, and a distance between the electrodes reduced to $D=\SIrange[range-phrase=-]{4}{8}{cm}$. In this experiment, the laser delivered \SI{30}{\fs} pulses of \SI{3}{\mJ}, with an initial beam diameter of \SI{1}{cm}. It generated one 20 cm-long filament. The occurrence, rate, and guiding of electrical discharges was characterized over periods of typically \SI{10}{s} (approximately 100 sparks in the free-running regime) on a standard video camera (Nikon D810).  

\section{Results and discussion}
\subsection{Low electric field (spark-free unloading)}

\begin{figure}[tb!]
\centering
\includegraphics[width=0.7\columnwidth, keepaspectratio]{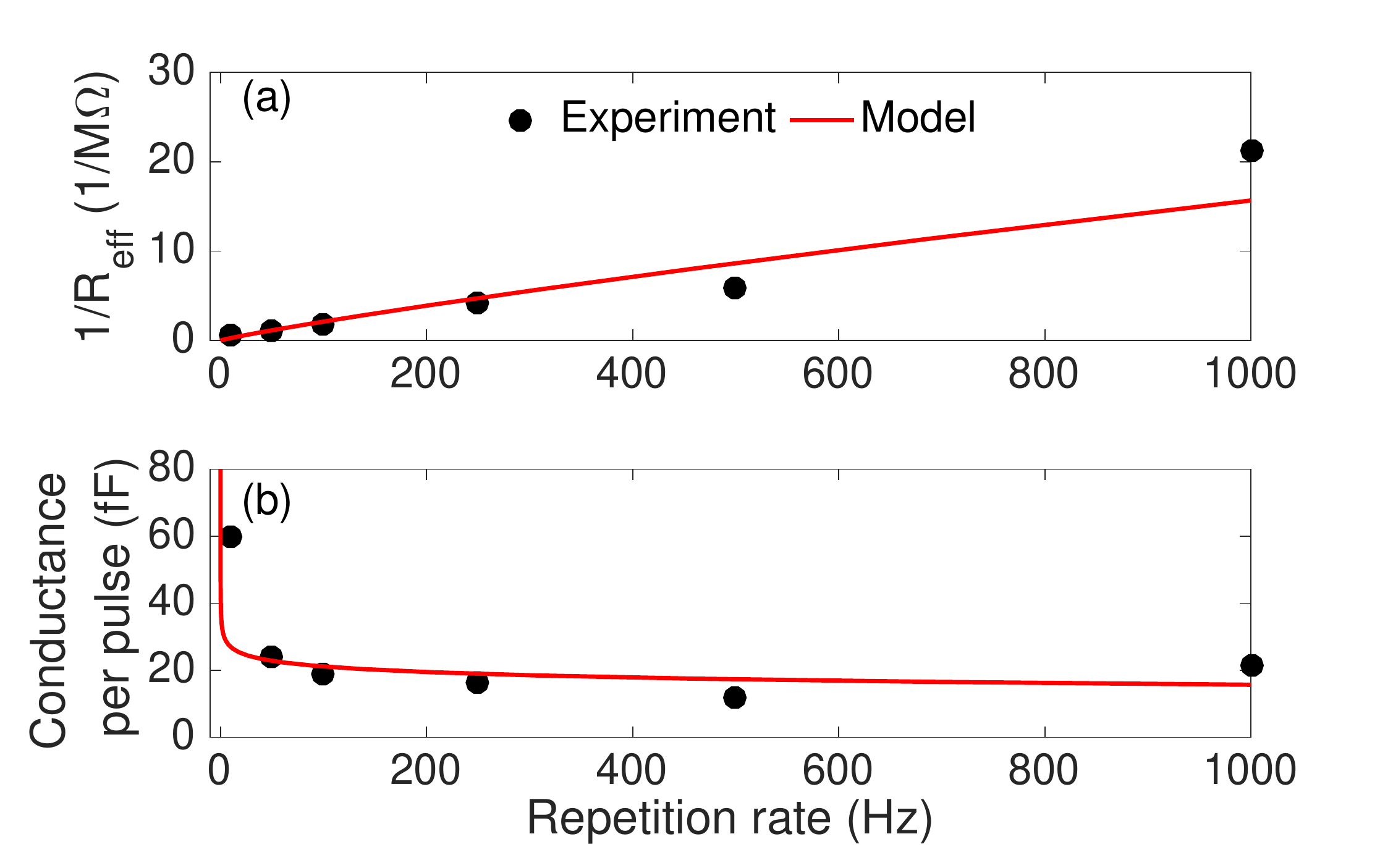}
\caption{Experimental and simulated repetition-rate dependence of spark-free unloading. (a) Effective conductance of the plasma channel; (b) \jk{conductance per pulse} between the electrodes separated by $D=\SI{16}{\cm}$ from each other, and $L = \SI{0.5}{\cm}$ from the laser filament.}
\label{fig:hvrep_repRate}
\end{figure}

The spark-free unloading of a charged capacitor has been observed at repetition rates down to \SI{10}{\Hz}~(\cref{fig:hvrep_repRate}). Higher repetition rates increase the effective conductance of the beam (\cref{fig:hvrep_repRate}a), although it reduces the \jk{conductance per} pulse (\cref{fig:hvrep_repRate}b): pulses \jk{unload the capacitor more efficiently} if they are separated by a longer time interval. 

Furthermore, \jk{we observed that} the efficiency of spark-free unloading is insensitive to wind blowing at \SI{2}{m/s} from the filaments towards the electrodes (blue fan in \cref{fig:hvrep_setup}). \jk{At \SI{2}{m/s}}, the displacement of the air mass between two pulses is \SI{2}{mm} at \SI{1}{kHz} repetition rate, much larger than the \SI{100}{\micro\m} diameter of the filaments. This excludes any direct interaction between successive laser pulses under wind. \jk{As the unloading is not affected by wind, we can conclude that it does not rely on interactions between successive pulses, even without wind.}

\jk{This, together with the} decay of the efficiency per pulse with increasing repetition rates seem to contradict the observation of thermal effects increasing the discharge triggering efficiency for repetition rates in the kilohertz range~\cite{Houard2016}. However, this apparent contradiction can be resolved by considering a toy model describing the temporal evolution of the conductivity between the electrodes (red curves in \cref{fig:hvrep_repRate}). This conductivity relies on ions flowing along the filament~\cite{schubert2015remote,Schubert2017} at an instantaneous current 
\begin{equation}
I = \frac{S \mu_{\textrm{ions}} V e N_{\textrm{ions}}}{D},
\end{equation}
with $N_{\textrm{ions}}$ the density of ions, $\mu_{\textrm{ions}}$ their mobility, $S$ the cross-section of the conducting region, $V$ the applied voltage, $D$ the distance between the electrodes, and $e$ the elementary charge. During a time $T$, the filament therefore conducts a charge 
\begin{equation}
Q = \frac{\mu_{\textrm{ions}} V e N_{\textrm{ions}}}{D} \int_0^T{S(t)N_{\textrm{ions}}(t) dt}.
\end{equation}

The evolution of the ionized cross section $S$ is driven by two processes. At microsecond-range times, the shockwave triggered by the energy (in the \si{\micro\J\per\cm}~\cite{Rosenthal2016}) deposited in the air by ionization~\cite{vidal2000modeling,Lahav2014,jhajj2014demonstration} sweeps the air around it at a speed of \SI{333}{m/s}~\cite{Lahav2014} up to a radius of \SI{740}{\micro\m}, diluting the ions over this region. At longer times, the ions are further diluted by diffusion, at a rate of $D_\textrm{d}=\SI{2e-5}{m^2/s}$ over a diameter slowly reaching \SI{3.4}{mm} after \SI{1}{s}. 

The decay time of the filament-induced ions is mainly governed by ion-ion recombination, which is a second-order kinetics: 
\begin{equation}
\frac{\mathrm{d} N_{\textrm{ions}}}{\mathrm{d}t} = -\beta N_{\textrm{ions}}^2, \label{eq:dNions}
\end{equation}
with $\beta=\SI{2e-13}{m^3/s}$~\cite{zhao1995femtosecond}. Therefore, the increase of the ionized cross section, by diluting the ions, drastically reduces their loss rate, and therefore increases the lineic ion concentration (hence, the resulting conductivity).

\begin{figure}[tb]
\includegraphics[width=0.7\columnwidth, keepaspectratio]{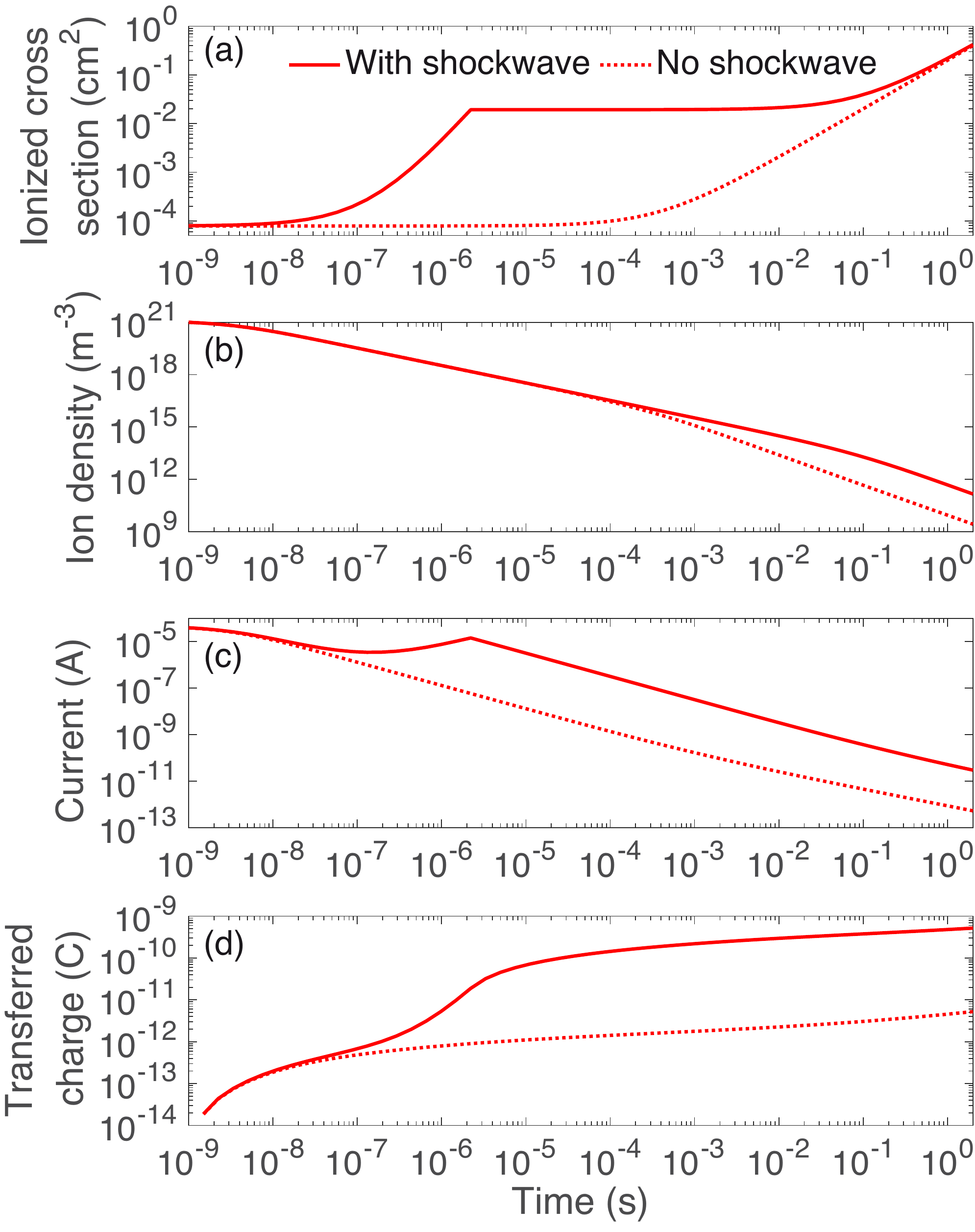}
\caption{Simulated \jk{temporal} evolution of (a) the ionized cross section, (b) ion density, (c) current, and (d) \jk{total} transported charge \jk{after a laser pulse}, with and without consideration of ion dilution due to the laser-induced shockwave.}
 	\label{fig:simul_evolution_shockwave}
\end{figure}

More specifically, the evolution of the ion density under the simultaneous influence of recombination and dilution expresses as 
\begin{equation}
\frac{\textrm{d}N_{\textrm{ions}}}{\textrm{d}t} = -\beta N_{\textrm{ions}}^2 - \frac{N_{\textrm{ions}}}{S} \frac{\textrm{d}S}{\textrm{d}t}. 
\end{equation}

Numerically integrating this evolution \jk{after a laser pulse occurring at $t=0$} (\cref{fig:simul_evolution_shockwave}) illustrates this drastic increase. The larger ionized cross-section between several tens of microseconds and \SI{1}{ms} (panel a) allows to keep a 4 times higher ion density (panel b), as compared with the simulation disregarding dilution. Convolved with the larger cross-section (panel a), this results in a 40 times lower resistance, hence a 40 times higher \jk{current} beyond \SI{2}{\micro\s} and up to at least \SI{100}{ms} (panel c). Integrating this intensity \jk{starting from the laser pulse ($t=0$)} results in a 70 times larger charge transfer (panel d). This total transferred charge still increases after \SI{1}{s}. \jk{When the repetition rate is increased, the time delay between two pulses is reduced. Therefore, each pulse re-ionises the medium before the conductivity has fully decayed. The previous pulse will then have transferred less charge than he would have if it were alone. On the other hand, due to the nonlinearity of \Cref{eq:dNions} and the associated fast initial decay of the ion density, the residual ions from the previous pulse have a negligible influence, preventing any cumulative effect. This explains the decaying unloading efficiency per pulse with increasing} laser repetition rate (red curves in \cref{fig:hvrep_repRate}).

The simulation \jk{considers} no direct interaction between successive pulses, consistent with the absence of influence of the wind. Rather, one can expect that wind displaces the conducting channel as a whole towards the ionized region close to the electrodes without substantially affecting its evolution. Furthermore, the ionized region extends over \SIrange{5}{10}{mm}, as visible by naked eye as a glow discharge. Therefore, displacements by a few millimeters of the plasma channels created by the laser filaments have little if any impact on the conduction. Note that the initial ion velocity is too small for its distribution to influence the width \jk{of the conducting channel}.

\subsection{Medium and high electric field: Spark guiding and suppression}

\begin{figure}[t]
\centering
\includegraphics[width=0.7\columnwidth, keepaspectratio]{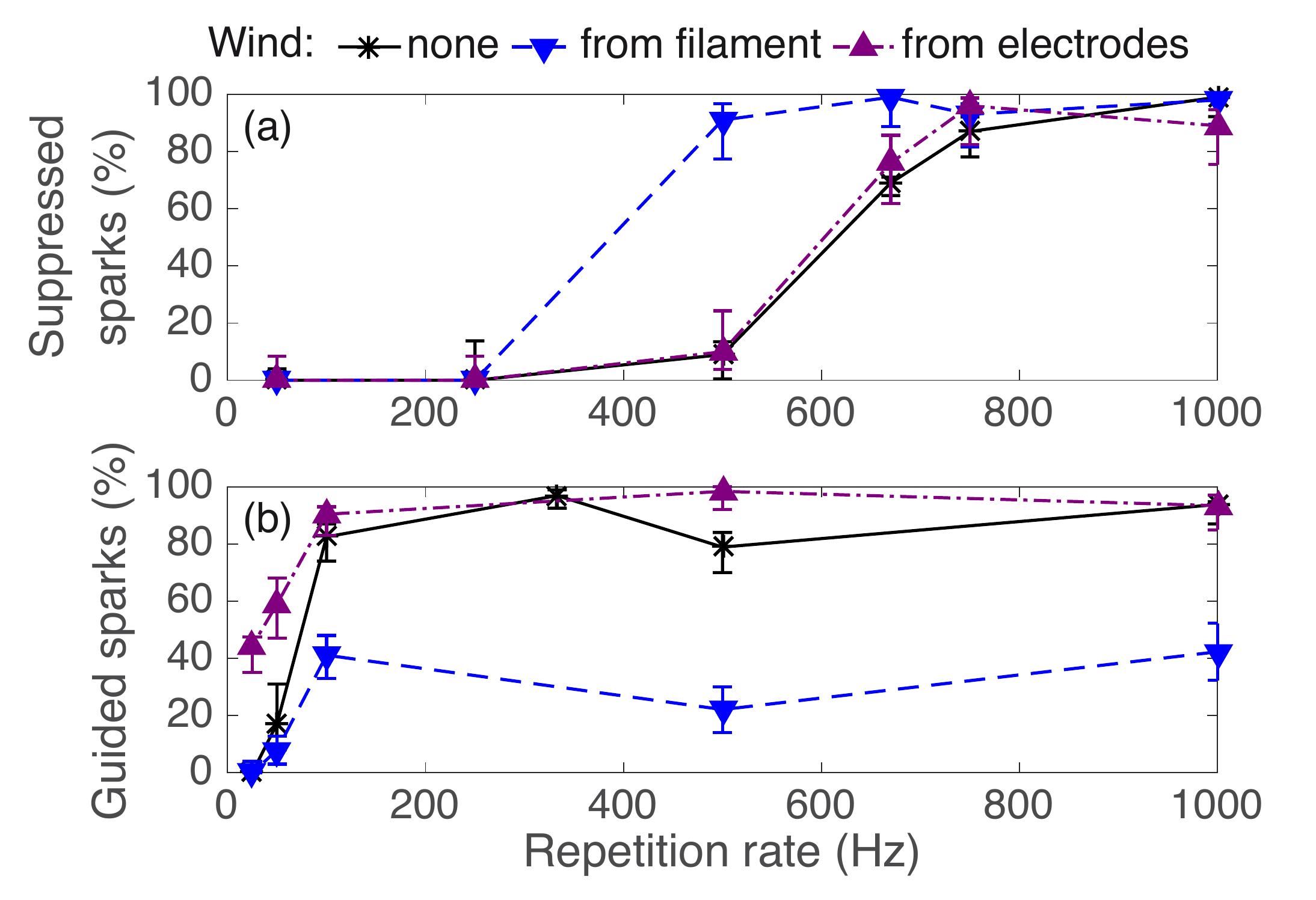}
\caption{Impact of wind and laser repetition rate on (a) spark suppression \jk{in a \SI{8}{cm} gap} and (b) guiding by laser filaments \jk{in a \SI{4}{cm} gap}. The blue and purple curves correspond to the fan positions with the same color in \Cref{fig:hvcapa_schema}. Error bars are evaluated by using the binomial law. }
 	\label{fig:effet_f_vent}
\end{figure} 

\Cref{fig:effet_f_vent} displays the effect of the laser on the \SI{100}{kV} discharges for electrodes separated from each other by \SI{8}{cm} (mean electric field of \SI{1.3e6}{V/m}, panel (a)) and \SI{4}{cm} (\SI{2.5e6}{V/m}, panel (b)), respectively. For the \SI{8}{cm} gap, the electric field is close to natural breakdown. As a consequence, by providing an extra conducting channel, the laser filaments pull more current from the power supply, resulting in a voltage drop that prevents the breakdown. Almost no spark is observed at a repetition rate of \SI{1}{kHz}. Reducing the repetition rate diminishes the effect of the laser, that stops influencing the discharges below \SI{500}{Hz} (\cref{fig:effet_f_vent}a).

The effect of the laser is reinforced by blowing wind at \SI{2}{m/s} from the filaments towards the electrodes (blue fan on \Cref{fig:hvrep_setup} and blue dashed curve on \Cref{fig:effet_f_vent}a), that decreases by \SI{200}{Hz} the laser repetition rate required to suppress the discharges. 
This can be understood as wind blowing the conducting channel towards the electrodes (by some millimeters in several milliseconds) favors its connection to the electrodes, hence the current flow and the voltage drop, resulting in a more efficient spark suppression.
Conversely, wind in the opposite direction (purple fan in \cref{fig:hvrep_setup} and purple dashed-dotted curve on \Cref{fig:effet_f_vent}a), parallel to the filaments, or blowing perpendicular to the filament-electrode plane has no influence on the effect of the laser. This is probably due to the fact that it degrades only slightly the already loose connection of the filament to the electrodes.
This behavior contrasts with the observation that, without laser, wind  from any direction reduces the rate of sparks by \SI{20}{\%}, with a maximal efficiency when blowing from the ground to the (positive) HV electrode, consistent with a conduction mainly carried by positive charge carriers, i.e., by ions. 

In the shorter gap (\SI{4}{cm}, \cref{fig:effet_f_vent}b), the laser-generated conducting channel is not sufficient  to let the voltage drop below the breakdown threshold. Therefore, discharges are not suppressed, but rather guided by the conducting filaments that offer a more favorable way to the discharges~\cite{ComtoCDGJJKFMMPRVCMPBG2000,LaCCDGJJKMMPRVPCM2000,rodriguez2002triggering,gordon2003streamerless,zhao1995femtosecond,Bazelyan2000,rambo2001high,IoninKLSSSSUZ2012}. Almost all discharges are guided for repetition rates above \SI{100}{Hz}, most probably related to the ability of the high-repetition rate beam to timely intercept streamer discharges before they develop into free-propagating sparks. 

Again, the wind has an asymmetric influence. However, wind blowing from the filament towards the electrodes (blue fan on \cref{fig:hvrep_setup} and blue dashed curve in \cref{fig:effet_f_vent}b), that was favorable to spark suppression, degrades discharge guiding. Conversely, when blowing from the electrodes towards the filament (purple fan on \cref{fig:hvrep_setup} and purple dash-dotted curve in \cref{fig:effet_f_vent}b), it tends to increase the fraction of guided discharges, especially at low repetition rates. 

As ions produced by the corona close to the electrodes are segregated by polarity, their diffusion-limited recombination ensures long lifetimes as compared with the laser repetition rate. Blowing the continuously produced ion cloud towards the beam position ensures immediate conduction when the filament is fired, allowing the discharge to follow this longer although easier guided way. Conversely, blowing them in the opposite direction reduces the guiding efficiency.

The contrast between the spark-free unloading in a low electric field, where the wind has no influence, and the stronger electric fields, where it can enhance or reduce the effect of the laser on the sparks, can be understood as follows. In the latter case, the effect of the laser critically depends on its electric connection to the electrodes. Therefore, displacing the plasma string produced by the filament with respect to them strongly influences the effect of the laser. Conversely, in the low-field unloading regime, the more diffuse effect reduces the impact of the plasma filament position.

Finally, let us note that the lateral displacement of long-lived, air depleted and ionized conducting channels generated by the laser filaments is reminiscent of multiple lightning strokes~\cite{Rakov2003}, also known as reilluminations or curtain lightning. In this natural process, an original lightning strike is followed by subsequent flashes propagating along the same path, that has moved as a whole with the air mass under the influence of wind. The delay between reilluminations ranges from tens to hundreds of milliseconds~\cite{Salimi2013,Mazur2016,Tarabukina2016,Wang2017,Johari2017} strikes, comparable with the conductivity lifetime resulting from shockwave-assisted discharges in our experiments.

\section{Conclusion}
In brief, for electric fields well below, close to, and above the threshold for discharges, ultrashort laser filaments induce remote spark-free unloading, discharge suppression, and discharge guiding, respectively. These effects rely on an indirect consequence of the thermal deposition, namely the dilution of the ions by the shockwave triggered by each laser shot. By quickly diluting the ions, this shockwave drastically limits recombination and maintains conductivity at measurable levels over tens to hundreds of milliseconds. Due to this single-shot, robust process, the influence of wind is mainly limited to displacing the ionized plasma channel created by the laser, influencing its connection with the electrodes. Therefore, the effect of wind depends on its direction, and is negligible when connection issues are not critical.  
In contrast, cumulative thermal effects, that require hundreds of laser pulses to be effective~\cite{Lahav2014,jhajj2014demonstration}, are not dominant, nor even required to explain the effect of the laser. 

\textbf{Acknowledgments}. 
This work was financially supported by the ERC advanced grant "Filatmo" and the European Union's Horizon 2020 research and innovation programme under grant agreement No 737033-LLR.
We gratefully acknowledge technical support by M. Moret and fruitful discussions with L. de la Cruz.

\bibliography{biblio_thesis}

\end{document}